\documentclass[aps,prb,twocolumn,showpacs,superscriptaddress]{revtex4-1}

\usepackage{amsmath}
\usepackage{amssymb}
\usepackage{graphicx}
\usepackage{dcolumn}
\usepackage{bm}
\usepackage{color}
\usepackage[usenames,dvipsnames]{xcolor}

\usepackage{bm}
\usepackage{graphicx}
\usepackage{ulem}

\def\be{\begin{equation}}
\def\ee{\end{equation}}
\def\ber{\begin{eqnarray}}
\def\eer{\end{eqnarray}}
\def\bwt{\begin{widetext}}
\def\ewt{\end{widetext}}

\begin{document}

\title {Electron polarization function and plasmons in metallic armchair graphene nanoribbons}
\author{A. A. Shylau}
\affiliation{Center for Nanostructured Graphene (CNG), Department of Micro and Nanotechnology,
Technical University of Denmark, DK-2800 Kongens Lyngby, Denmark}
\author{S. M. Badalyan}
\email{sambad@nanotech.dtu.dk}
\affiliation{Center for Nanostructured Graphene (CNG), Department of Micro and Nanotechnology,
Technical University of Denmark, DK-2800 Kongens Lyngby, Denmark}
\affiliation{Department of Physics, University of Antwerp, Groenenborgerlaan 171, B-2020 Antwerpen, Belgium}
\author{F. M. Peeters}
\affiliation{Department of Physics, University of Antwerp, Groenenborgerlaan 171, B-2020 Antwerpen, Belgium}
\author{A. P. Jauho}
\affiliation{Center for Nanostructured Graphene (CNG), Department of Micro and Nanotechnology,
Technical University of Denmark, DK-2800 Kongens Lyngby, Denmark}
\date{\today}

\begin{abstract}
Plasmon excitations in metallic armchair graphene nanoribbons are investigated using the random phase approximation. An exact analytical expression for the polarization function of Dirac fermions is obtained valid for arbitrary temperature and doping. We find that at finite temperatures, due to the phase space redistribution among inter-band and intra-band electronic transitions in the conduction and valence bands, the full polarization function becomes independent of temperature and position of the chemical potential. It is shown that for a given width of nanoribbon there exists a single plasmon mode whose energy dispersion is determined by the graphene's fine structure constant. In the case of two Coulomb-coupled nanoribbons, this plasmon splits into in-phase and out-of-phase plasmon modes with splitting energy determined by the inter-ribbon spacing.
\end{abstract}


\maketitle

\section{Introduction}

Plasmons, i.e. collective charge density oscillations of free carriers, have attracted considerable fundamental~\cite{Pines1966,GV2005} and practical~\cite{Maier2007} interest for several decades. Plasmons in graphene~\cite{Geim2004,GeimNovo2007} exhibit a number of extraordinary new features such as a high degree of optical field confinement~\cite{Geim2008,Jablan2009,Koppens2012} and gate-tunability~\cite{Vakil2011,Fang2012,Yan2012}, which has a potential for practical applications. Particularly, achieving electric control of light is one of the key challenges to efficient plasmonic technology~\cite{Polini2012}.

Recently plasmons in graphene~\cite{Stauber2014} have been studied in single~\cite{Vafek2006,Guinea2006,Sarma2007,Barlas2007,Abedinpour2011,Li2013,Fei2013,Alonso2014} and spatially separated double~\cite{Sarma2009,Stauber2012,Profumo2012,SMB2012,SMB2012R,Shylau2014} and multilayer~\cite{Jhu2013} structures. It was shown that finite temperature can strongly affect the plasmon modes~\cite{Jhu2013,SMB2012,Vafek2006,Li2013}.  Particularly, at zero temperature in Coulomb-coupled $N$-layer  unbalanced graphene structures there exists only one in-phase optical plasmon mode and $N-N_0-1$ out-of-phase acoustical modes~\cite{SMB2012,Jhu2013,Sarma2009} (here $N_0$ is the number of neutral graphene layers in such multilayer structures). On the other hand, at finite temperatures~\cite{Jhu2013,SMB2012}, due to thermally activated electronic transitions, these $N_0$ passive acoustical modes become separated from the top of the electron-hole continuum and at temperatures of the order of the Fermi energy they behave completely similar to out-of-phase 
plasmon modes in multilayer structures with equal finite carrier densities in each layer. Intrinsic thermal plasmon modes are formed also in single-layer neutral graphene structures~\cite{Vafek2006,Li2013}. Importantly, in the long wavelength limit these temperature induced in-phase and out-of-phase intrinsic plasmon modes in multilayer graphene structures preserve their characteristic square-root and linear energy dispersions, are Landau damped, and disappear at vanishing temperatures.

The situation is qualitatively different in graphene nanoribbons, specifically in armchair metallic nanoribbons with a single-particle Dirac spectrum of massless chiral fermions~\cite{BF2006}.  As shown in Ref.~\onlinecite{BF2007} when the chemical potential is at the charge neutrality point there exist intrinsic plasmons at zero temperature and these plasmons are not Landau damped.

\begin{figure}[t]
\includegraphics[width=\columnwidth]{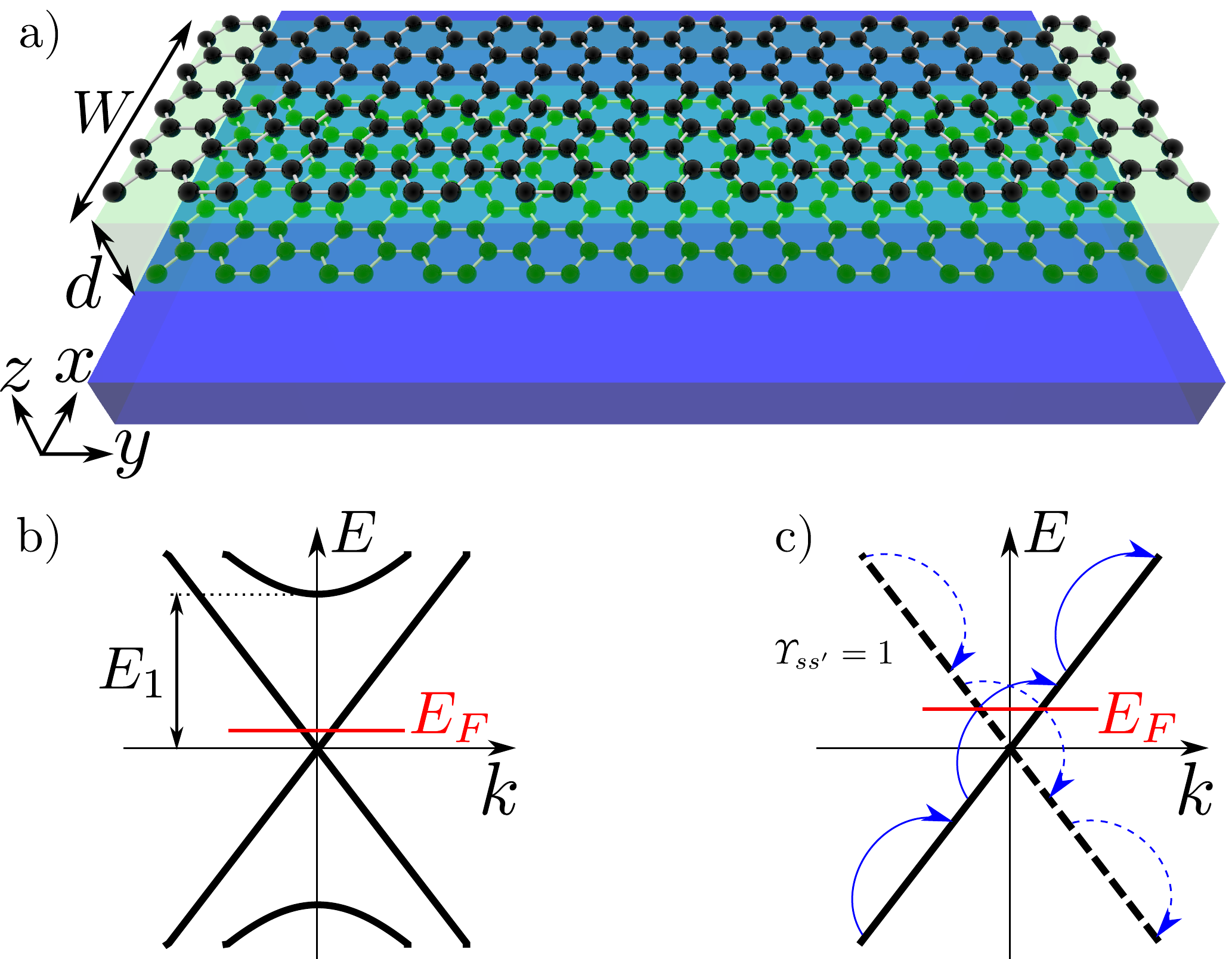}
\caption{(Color online) Schematic representation of two parallel metallic armchair graphene nanoribbons of width $W$ and inter-ribbon spacing $d$ (a). The single-electron energy spectrum (b) and its zoomed in (c) for a close-up for energies $E \ll E_1$.  The blue solid and dashed arrows in (c) depict various inter-band and intra-band electronic transitions for a given bosonic momentum $q$ at $T\neq0$ and $E_{\text{F}}\neq0$. There are two different transition channels along the solid and dashed lines, for which the spinor form factors $\varUpsilon_{s,s'}\neq 0$.}
\label{fig1}
\end{figure}

In this article we study the effect of finite temperature and doping on the plasmon modes in individual and Coulomb-coupled metallic armchair graphene nanoribbons, schematically shown in Fig.~\ref{fig1}. First we calculate the Lindhard polarization function using the standard approach based on the random phase approximation~\cite{GV2005}. The previous works on this system neglect either the intra-band contribution to the finite temperature polarization function~\cite{BF2007} or the inter-band contribution to the zero temperature polarization function~\cite{Hai2013,Bahrami2014}. Here, we fully take into account both the inter-band and intra-band parts of the polarization function for arbitrary temperature and position of the chemical potential, within the lowest subband of transverse quatization. We find that due to the chiral and one-dimensional nature of Dirac fermions, when the inter-band contribution is taken separately, it exhibits a strong temperature effect. Within the $0<q<q_T$ window of the bosonic momentum $q$, this contribution changes qualitatively its long wavelength 
behavior in comparison with that obtained at $T=0$. Here $q_T=T/v_\text{gr}$ and $T$ is the temperature and $v_\text{gr}$ the velocity of Dirac fermions in graphene. We use units with Planck's and Boltzmann's constants $\hbar=k_B=1$. We observe, however, that with increasing $T$ the phase space among the inter-band and intra-band electronic transitions is redistributed in such a way that the total phase space is determined only by the transition momentum $q$. As a result, the full polarization function becomes independent of the position of the chemical potential and temperature. 

We find that within the random phase approximation for a given width, $W$, of individual nanoribbons and for arbitrary values of $T$ and the Fermi energy, $E_\text{F}$, there exists a {\it single} plasmon mode with energy dispersion determined by the graphene's fine structure constant~\cite{Kotov2012}. This plasmon emerges with a logarithmically divergent velocity for vanishing $q W$ and remains Landau undamped even for large values of $q W$. In Coulomb-coupled double nanoribbons, this plasmon splits into in-phase and out-of-phase plasmon modes with splitting energy determined by the inter-ribbon spacing $d$. The out-of-phase plasmon shows a strictly linear energy dispersion in the long wavelength limit with a velocity increasing as $\sqrt{d}$. The velocity of the in-phase plasmon mode is logarithmically divergent for vanishing $q$ and is larger by a factor of $\sqrt{2}$ than the velocity of the plasmon in the individual nanoribbons.

\section{Theoretical model}

We obtain plasmon excitations from the poles of the exact Coulomb propagator, $\hat{V}(q,\omega)$, which satisfies the matrix Dyson equation
\begin{equation}
\hat{V}(q,\omega)=\hat{v}(q)+\hat{v}(q)\cdot\hat{\Pi}(q,\omega)\cdot\hat{V}(q,\omega)~.
\label{dyson}
\end{equation}
Here $q$ and $\omega$ are the bosonic momentum and frequency, $\hat{\Pi}(q,\omega)$ is the irreducible dynamical polarization function of electrons, and $\hat{v}(q)$ the bare Coulomb interaction. In structures consisting of two subsystems $\hat{v}(q)\equiv v^{ij}(q)$ is a tensor with respect to the subsystem indices $i,j=1,2$ and its three different components, $v^{11}(q)$, $v^{22}(q)$, and $v^{12}(q)=v^{21}(q)$, describe the intra- and inter-subsystem interactions, which determine the properties of the exact Coulomb propagator. In graphene nanoribbons the matrix elements of bare interactions $v^{ij}(q)$ depend additionally on the quantum indices $n$ and $s$, corresponding to the restricted motion of electrons along the transverse $x$ direction and  the chirality of Dirac fermions, respectively. We assume that the spacing $d$ between armchair graphene nanoribbons is along the $z$ direction so that the carrier density in both nanoribbons can be written as $\rho(z)= \delta\left(z-d \left(1-\delta _{i j }\right)
\right)$, using the Dirac and Kronecker delta functions. The electron wave functions in the $(x,y)$ plane are~\cite{BF2006}
\begin{equation}
\Psi_{n s}(x,y)=\frac{e^{i k_y y}}{2 \sqrt{\frac{a_0}{2}+W} \sqrt{L}}
\left(
\begin{array}{c}
 e^{i x k_n-i \theta \left(k_n,k_y\right)} \\
s e^{i x k_n}\\
\begin{array}{c}
 -e^{-i k_n x -i \theta \left(k_n,k_y\right)} \\
 s e^{-i k_n x}
\end{array}
\end{array}
\right)
\label{wf}
\end{equation}
where $L$ is a normalization length along the $y$ direction, $W$ the width of the nanoribbon along the $x$ direction,  $a_{0}$ the lattice parameter of graphene, and $\theta(k_{n}, k_{y}) = \arctan(k_{n}/k_{y})$. The transverse quantization subband index is an integer, $n=0,\pm 1, \pm2,\dots$ and the chiral index $s$ describes the conduction ($s = +1$) and the valence ($s = -1$) bands. The single-particle  energy of electrons is $E_{ns}(k_{y}) = s v_{\text{gr}}\sqrt{k_{n}^{2} + k_{y}^{2}}$ with the transverse momentum $k_{n}=\frac{\pi  n}{W+a_0/2}+\frac{2 \pi }{3 a_0}$ along the $x$ direction, and the conserved momentum $k_{y}$, corresponding to the translational invariance of the nanoribbon, is along the $y$ direction.  For a special choice of the width of the nanoribbon $W= (3 M_{0}+1)a_0 $, we have $k_{n}=\frac{2 \pi  (n+2M_{0}+1)}{3a_0 (2 M_{0}+1)}$ where  $2 M_{0} + 1 + n$ is an integer too. Hence, for any value of $W$ there exists an $n$ for which the armchair nanoribbon has a Dirac spectrum with $E_{ns}(k_
{y}) = s v_{\text{gr}}k_{y}$, i.e., the nanoribbon is metallic.
Further, we assume that the Fermi energy, $E_{\text{F}}$, and the temperature satisfy $E_{\text{F}}, T\ll v_{\text{gr}}k_{1}$, so that electronic transitions between the transverse subbands do not contribute significantly to the polarizability.

\begin{figure}[t]
\includegraphics[width=.9\columnwidth]{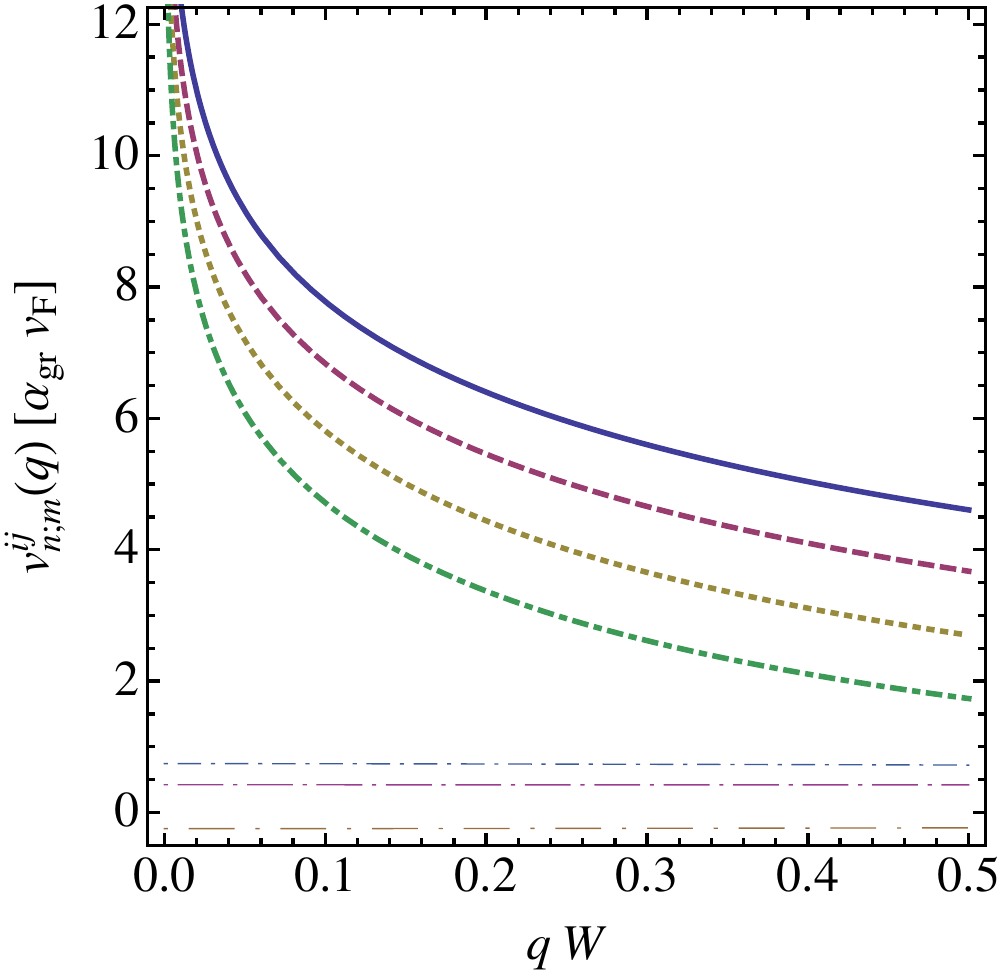}
\caption{(Color online) The Coulomb matrix elements in double armchair nanoribbons. The thick solid, dashed, dotted, and dot-dashed curves correspond, respectively, to $v_{0;0}^{12}(q,d,W)$ for spacing $d/W=0$, $0.2$, $0.5$, and $1$. The thin curves correspond to $v_{1;1}^{11,22}(q,0,W)$, $v_{2;2}^{11,22}(q,0,W)$, and $v_{2;0}^{11,22}(q,0,W)$.}
\label{fig2}
\end{figure}

The poles of the exact Coulomb propagator (\ref{dyson}) are given by the zeros of the determinant
\begin{equation}
\text{det} \left| \delta^{ij}\delta_{\alpha,\alpha'}\delta_{\beta,\beta'}-v^{ij}_{\alpha \beta;\alpha' \beta'}(q,d,W)
\Pi^{j}_{\alpha' \beta'}(q,\omega) \right|=0
\label{det}
\end{equation}
where the Greek letters denote the combined quantum numbers $\alpha=\{n,s\}$. The matrix elements of intra- and inter-ribbon Coulomb interaction are
\begin{eqnarray}
v_{\alpha\beta; \alpha' \beta'}^{i j}(q,d,W)&=&\int^1_0 dx dx' \psi _{\alpha}^{*}(x) \psi _{\beta }^{*}(x') \\
&\times& \bar{v}^{i j}(q,d,W)\psi_{\alpha'}(x) \psi_{\beta'}(x') \nonumber
\label{cme}
\end{eqnarray}
where $\bar{v}^{i j}(q,d,W)= 2 \alpha_{\text{gr}} v_{\text{gr}}K_0 (q [d^2 \left(1-\delta _{i j}\right)+W^2\left(x-x'\right)^2]^{1/2})$ with $\alpha_{\text{gr}}=\frac{e^2}{v_{\text{gr}}\epsilon_{\text{eff}}}\approx \frac{2.2}{\epsilon_{\text{eff}}}$ the fine structure constant of graphene, embedded in a dielectric medium with an effective low frequency dielectric function $\epsilon_{\text{eff}}$, and $K_{0}$ is the modified Bessel function of the second kind. Making use of the armchair nanoribbon wave functions (\ref{wf}), the Coulomb matrix elements become~\cite{BF2007}
\begin{eqnarray}
v_{n-m; n'-m'}^{i j}(q,d,W)&=&\int_{0}^{1} dx dx' \cos \left(\pi  x \left( n - m \right)\right)   \\ \nonumber
&\times& \cos \left(\pi  x'  \left(n' - m' \right)\right) \bar{v}^{i j}(q,d,W)~,
\label{cme2}
\end{eqnarray}
which are independent of the chiral indices. As seen from Fig.~\ref{fig2} the intra- and inter-ribbon Coulomb matrix elements $v_{0; 0}^{i j}(q,d,W)$ with $n-m=n'-m'=0$, corresponding to the electronic transitions within the same transverse subband, are larger than the other elements as long as $q$ is not too large, and hence dominate  the determinant in Eq.~(\ref{det}). Therefore, as a first approximation the determinant can be reduced to the dielectric function of double nanoribbons,
\begin{equation}
\varepsilon(q,\omega)=\varepsilon_{1}(q,\omega)\varepsilon_{2}(q,\omega)-v^{12}_{0;0}(q,d,W)^{2}\Pi^{1}(q,\omega)\Pi^{2}(q,\omega)
\label{screening}
\end{equation}
with
\begin{equation}
\varepsilon_{1,2}(q,\omega)=1-v^{11}_{0;0}(q,0,W)\Pi^{1,2}(q,\omega)~.
\label{singscreening}
\end{equation}
Here $\varepsilon_{1,2}(q,\omega)$ and $\Pi^{1,2}(q,\omega)=\sum_{n}\Pi^{1,2}_{nn}(q,\omega)$ are the dynamical dielectric function and the dynamical polarization function in each graphene nanoribbon, respectively.

\section{Polarization function in metallic armchair graphene nanoribbons}

In this section we calculate the Lindhard polarization function of individual metallic armchair graphene nanoribbons for an arbitrary position of the chemical potential and temperature.
The polarization function is calculated within the random phase approximation taking into account electronic transitions within and between the lowest conduction band and the highest valence band ({\it cf.} Fig.~\ref{fig1}(c)). As discussed above the contribution from other subbands is small. No other approximation is made in the calculation.

The polarization function is defined as a sum over the chirality indices as $\Pi^{i}(q,\omega )=\sum_{s,s'=\pm1}\Pi_{s,s'}^{i}(q,\omega )$ with
\begin{eqnarray}
\Pi_{s,s'}(q,\omega )&=&\frac{g_s}{L} \sum_{k_y}  
\frac{f_T\left(E_{s'}\left(k_y+q\right)\right)-f_T\left(E_{s}\left(k_y\right)\right)}
{E_{s'}\left(k_y+q\right)-E_{s}\left(k_y\right)-\omega-i0} \nonumber \\
&\times&\varUpsilon_{s,s'}\left(k_y,k_y+q\right)~.
\label{PF}
\end{eqnarray}
Here $g_s$ is the spin degeneracy factor, $f_T(E)$ the Fermi distribution function, and the spinor overlapping form factors are given by
\begin{eqnarray}
\varUpsilon_{s,s'}\left(k_y,k_y+q\right)=\frac{1}{2} \left(1+ s s' \cos\theta _q \right)
\label{soff}
\end{eqnarray}
with $\theta_q$ defined as an angle between the $(0,k_y)$ and $(0,k_y+q)$ electron momenta.

In general, the inter-band electronic transitions between the valence and conduction bands together with the intra-band transitions within each band contribute to the polarization function and below we consider them separately.

\subsection{Inter-band contribution to the polarization function}

In an individual metallic armchair graphene nanoribbon the contribution to the polarization function made by electronic transitions between the conduction and valence bands can be represented~\cite{BF2007} in the following way
\begin{eqnarray}
\Pi_{\text{inter}}(q,\omega)=
\frac{g_{s}}{\pi}\frac{v_{\text{gr}}q}{\omega^2-v_{\text{gr}}^{2} q^2} F_{\text{inter}}(q,E_\text{F},T)~.
\label{pfinter}
\end{eqnarray}
Here for the function $F_{\text{inter}}(q,E_\text{F},T)$, which is determined by the phase space of electron-hole fluctuations, we find for arbitrary values of $T$ and $E_\text{F}$~\cite{footnote}
\begin{eqnarray}
&&F_{\text{inter}}(q,E_\text{F},T)=-\frac{1}{v_{\text{gr}}}\int_{0}^{v_{\text{gr}}q}dE \left[f_{T}(E)-f_{T}(-E)\right]  \\
&&=-q-\frac{T}{v_{\text{gr}}}\ln \frac{\left(1+e^{-E_\text{F} /T}\right) 
\left(1+e^{E_\text{F} /T}\right)}
{\left(e^{v_{\text{gr}} q / T}+e^{-E_\text{F} / T}\right) 
\left(e^{v_{\text{gr}} q / T}+e^{ E_\text{F} / T}\right)}~.
\label{phsinter}
\end{eqnarray}
There are two relevant momentum scales, $q_T = T/v_\text{gr}$ and $1/W$, and the long wavelength limit here is defined as $q\ll q_T, 1/W$.
We observe that because of the chiral and one-dimensional nature of Dirac fermions, the order of limits $T\rightarrow$ and $q\rightarrow 0$ is important in determining the behavior of the function  $F_{\text{inter}}(q,E_\text{F},T)$ and this is independent of the position of the chemical potential. 

If $T <v_{\text{gr}}q \rightarrow 0$, we have from Eq.~(\ref{phsinter})
\begin{eqnarray}
F_{\text{inter}}(q,E_\text{F},T)\approx
\begin{cases}  q-E_\text{F}/v_{\text{gr}}  & \mbox{if } v_{\text{gr}} q > E_\text{F} \\
0 &\mbox{if } E_\text{F} > v_{\text{gr}} q~,
\end{cases} 
\label{phsinterT0}
\end{eqnarray}
i.e. at $T=0$ the phase space vanishes linearly in $q$ in neutral graphene nanoribbons. We note that in doped samples the inter-band contribution to the polarization function, which is finite for $v_{\text{gr}} q > E_\text{F}$, decreases with $E_\text{F}$. This term has not been considered in previous calculations at zero temperature~\cite{Hai2013,Bahrami2014}. Meanwhile one can see below that this contribution together with the respective term of the intra-band contribution in Eq.~(\ref{phsintraT0}) are responsible for the disappearance of the dependence of the full polarization function on $E_\text{F}$.

If $T>v_{\text{gr}}q \rightarrow 0$, we find from Eq.~(\ref{phsinter})
\begin{eqnarray}\label{phsinterq0}
F_{\text{inter}}(q,E_\text{F},T) \approx \frac{v_{\text{gr}}q^2}{4T} \left(1-\frac{E_\text{F}^2}{4 T^2}\right)~,
\end{eqnarray}
i.e. at finite temperatures the phase space vanishes much slower, i.e. quadratically in $q$. Thus, at finite temperatures a momentum window, $0 < q < q_T$, is opened where the behavior of the inter-band contributions to the polarization function becomes qualitatively different. In the limit of $T\rightarrow 0$, this momentum subregion disappears and the polarization function tends smoothly to that obtained for $T = 0$ in the limit of $q < 1/W$.

If we restrict ourselves to the contribution to the polarization function, made only by inter-band electronic transitions ({\it cf.} Ref.~\onlinecite{BF2007}), we can see in the Appendix that this subtle point related to the order of limits $T\rightarrow 0$ and $q\rightarrow 0$ can change qualitatively the plasmon energy dispersion at finite temperatures in comparison with that obtained at zero temperature~\cite{BF2007}. 

\subsection{Intra-band contribution to the polarization function}

At finite temperatures electronic transitions both within the conduction band and within the valence band contribute to the intra-band part of the polarization function,
\begin{eqnarray}
\Pi_{\text{intra}}(q,\omega)=
\frac{g_{s}}{\pi}\frac{v_{\text{gr}}q}{\omega^2-v_{\text{gr}}^{2} q^2} F_{\text{intra}}(q,E_\text{F},T)~.
\label{pfintra}
\end{eqnarray}
For arbitrary values of $T$ and $E_\text{F}$, we obtain
\begin{eqnarray}
&&F_{\text{intra}}(q,E_\text{F},T)= -\frac{1}{v_{\text{gr}}}\int_{0}^\infty dE \left(\left[f_{T}(E+v_{\text{gr}}q) \right. \right.  \\
&&\left. \left. - f_{T}(E)\right]  + \left[f_{T}(-E)-f_{T}(-E-v_{\text{gr}}q)\right]  \right) \nonumber
\\
&&=\frac{T}{v_{\text{gr}}}
\ln \frac{\left(1+e^{-E_\text{F} /T}\right) \left(1+e^{E_\text{F} /T}\right)}
{\left(e^{-v_{\text{gr}} q / T}+e^{-E_\text{F} / T}\right) \left(e^{-v_{\text{gr}} q / T}+e^{ E_\text{F} / T}\right)}~.
\label{phsintra}
\end{eqnarray}
The behavior of the function  $F_{\text{intra}}(q,E_\text{F},T)$ in the long-wavelength limit also depends on the order of the $q\rightarrow 0$ and $T\rightarrow 0$ limits. 

If $T <v_{\text{gr}}q \rightarrow 0$, we have
\begin{eqnarray}
F_{\text{intra}}(q,E_\text{F},T)\approx
\begin{cases} E_\text{F}/v_{\text{gr}}    & \mbox{if } v_{\text{gr}} q > E_\text{F} \\
q  &\mbox{if } E_\text{F} > v_{\text{gr}} q~.
\end{cases} 
\label{phsintraT0}
\end{eqnarray}
As expected, at zero temperature the scattering phase space vanishes identically in neutral graphene nanoribbons while it is linear in $q$ or a constant in doped samples. 

If $T>v_{\text{gr}}q \rightarrow 0$, we find
\begin{eqnarray}
F_{\text{intra}}(q,E_\text{F},T) \approx q - \frac{v_{\text{gr}}q^2}{4T} \left(1-\frac{E_\text{F}^2}{4 T^2}\right)~,
\label{phsintraq0}
\end{eqnarray}
i.e. in contrast to the inter-band contribution ({\it cf.} Eq.~(\ref{phsinterq0})), at finite temperatures the phase space for the intra-band transitions vanishes much faster, i.e. linearly in $q$. 

As seen, however, from Eqs.~(\ref{phsinterT0}) and (\ref{phsintraT0}) as well as from Eqs.~(\ref{phsinterq0}) and (\ref{phsintraq0}), for both $T=0$ and $T\neq 0$ the phase space is redistributed among the inter-band and the intra-band transitions in such a way that it cancels exactly the $q^0$ and $q^2$ dependence, inherent in the separate intra-band and inter-band contributions. Strikingly, we find that this cancelation takes place without the long wavelength limit restriction on $q$ and for arbitrary values of $T$ and $E_\text{F}$ so that the total phase space is
\begin{eqnarray}
F(q)\equiv F_{\text{inter}}(q,E_\text{F},T)+F_{\text{intra}}(q,E_\text{F},T) = q~.
\label{phstot}
\end{eqnarray}
Thus, the full Lindhard polarization function in metallic armchair graphene nanoribbons is given by the following exact formula
\begin{eqnarray}
\Pi(q,\omega)=\frac{g_{s}}{\pi}\frac{v_{\text{gr}}q^2}{\omega^2-v_{\text{gr}}^{2} q^2} 
\label{pffinal}
\end{eqnarray}
and, in contrast to the previous findings~\cite{BF2007,Hai2013,Bahrami2014}, it shows a universal behavior, independent of the position of the chemical potential and the value of temperature. 

\begin{figure}[t]
\includegraphics[width=.9\columnwidth]{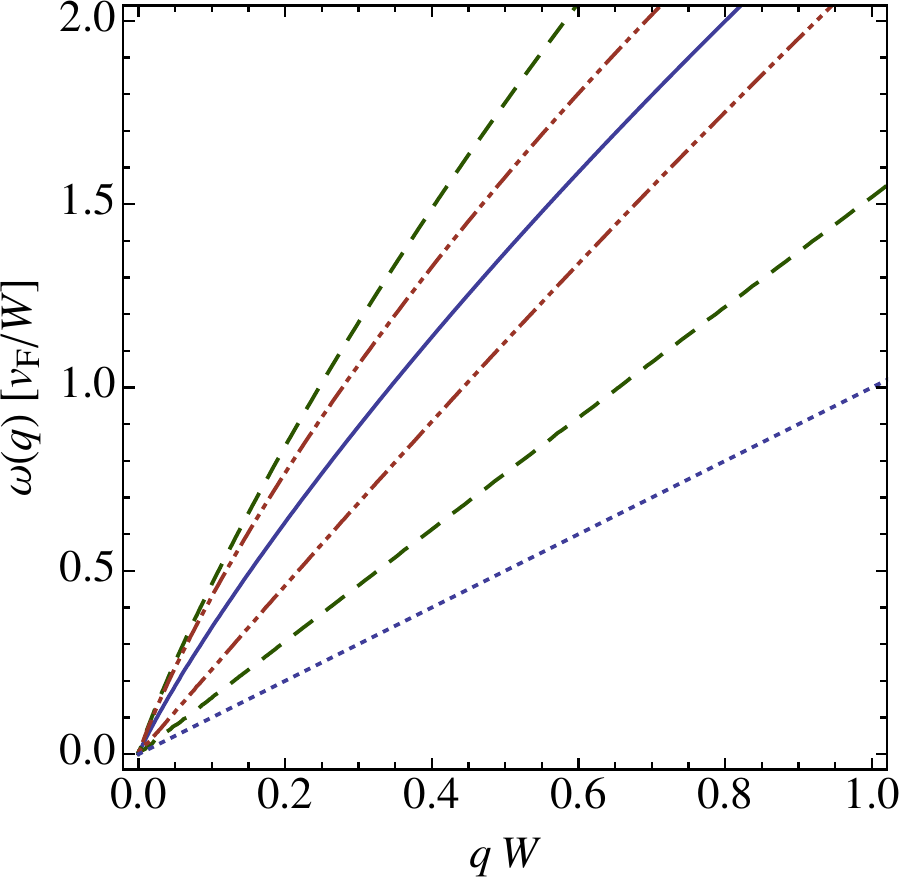}
\caption{(Color online) Plasmon dispersions in single and double structures of metallic armchair graphene nanoribbons. The solid curve represents the plasmon mode in individual nanoribbons. The upper and lower dashed  (dash-dotted) curves correspond, respectively, to the in-phase and out-of-phase plasmon modes for spacing $d=0.2W$ ($d=W$). The dotted line is the single-particle Dirac spectrum.}
\label{fig3}
\end{figure}

\section{Plasmons in metallic armchair graphene nanoribbon structures}

In individual graphene nanoribbons we obtain the plasmon dispersion from Eq.~(\ref{singscreening}). In the long wavelength limit, $v_{\text{gr}}q/T \rightarrow 0$, the polarization function (\ref{pffinal}) is approximated as
\begin{equation}
\Pi(q,\omega)\approx\frac{2}{\pi}\frac{v_{\text{gr}} q^2}{\omega^2}~.
\label{piflwll}
\end{equation}
In this limit making use of the asymptote of the Coulomb matrix element $v^{11}_{0;0}(q,0,W)\approx -2 \alpha_{gr} v_{\text{gr}}\ln{(q W)}$, we find the plasmon dispersion $\omega_{p}(q) = c_{p}(q) q$ with its velocity given by
\begin{equation}
c_{p}(q) \approx 2\left(-\frac{\alpha_{\text{gr}}}{\pi} \ln(q W) \right)^{1/2}v_{\text{gr}}~.
\label{enlwll}
\end{equation}
It is seen that the plasmon mode emerges from the origin of $(\omega,q)$ plane with a logarithmically divergent velocity, $c_{p}(q)\propto \sqrt{-\ln(q W)}$.


In Coulomb-coupled graphene structures, consisting of two spatially separated metallic armchair nanoribbons with arbitrary doping levels, we represent the dielectric function (\ref{screening}) as $\varepsilon(q,\omega)=\varepsilon^{+}(q,d,W)\varepsilon^{-}(q,d,W)$ where
\begin{eqnarray}
\varepsilon^{\pm}(q,d,W)=1-v^{\pm}_{0;0}(q,d,W)\Pi(q,\omega)
\label{screening2}
\end{eqnarray}
with $v_{0;0}^{\pm}(q,d,W)=v_{0;0}^{11}(q,0,W)\pm v_{0;0}^{12}(q,d,W)$. In the limit of vanishing $q$, we approximate
\begin{eqnarray}
v_{0;0}^{+}(q,d,W)&\approx& -4\alpha_{gr} v_{\text{gr}}\ln(q W)~,\\
v^{-}_{0;0}(q,d,W) &\approx& 2\alpha_{gr} v_{\text{gr}}[
(d^2/W^2) \ln (d/W)\nonumber\\
&\quad&+(1/2)(1-d^2/W^2 \log (1+d^2/W^2)\nonumber\\
&\quad&+ 2(d/W) \cot^{-1}(d/W)]~.
\label{CMEDL}
\end{eqnarray}
Assuming $d\ll W$, we have $v^{-}_{0;0}(q,d,W)\approx 2\pi \alpha_{gr} v_{\text{gr}}d/W$.
Therefore, from Eqs.~(\ref{piflwll}) and (\ref{screening2})-(\ref{CMEDL}) for the dispersion relations of the in-phase and out-of-phase plasmon modes in the long wavelength limit, we find
\begin{eqnarray}
\omega_{+}(q\rightarrow 0)&\approx& 2 \left(-\frac{2\alpha_{\text{gr}}}{\pi} \ln(q W) \right)^{1/2}v_{\text{gr}}q \\
\omega_{-}(q\rightarrow 0)&\approx & 2 \left(\alpha_{\text{gr}} \frac{d}{W}  \right)^{1/2}v_{\text{gr}}q
\label{dlenlwll}
\end{eqnarray}
It is seen that the out-of-phase plasmon mode, $\omega_{-}(q)$, is strictly linear in the momentum $q$ and its velocity shows a square-root dependence on the inter-ribbon spacing, $c_{-}\propto \sqrt{d}$. The velocity of the in-phase plasmon mode, $c_{+}(q)$, diverges logarithmically for vanishing $q$. In this limit $c_{+}(q)$ is by a factor of $\sqrt{2}$ larger than the plasmon velocity $c_{p}(q)$ in individual nanoribbons and, in contrast to the acoustical mode, is independent of $d$.

In Fig.~\ref{fig3} we plot the energy dispersion of the plasmon in individual nanoribbons together with the in-phase and out-of-phase plasmon modes of double nanoribbons without the long wavelength restriction on $q$. As seen for both $d=0.2W$ and $d=W$  the energy of the in-phase plasmon mode is closer to the plasmon energy in individual graphene nanoribbons than the energy of the out-of-phase plasmon mode, which therefore increases faster with $d$. For large values of $d$ the Coulomb matrix elements $v^{\pm}_{0;0}(q)\rightarrow v^{11}_{0;0}(q)$ and the energy dispersions of the in-phase and out-of-phase modes tend to the plasmon energy in individualal graphene nanoribbons from above and below, respectively.

In our calculations we assume that structures of metallic armchair graphene nanoribbons are embedded in a homogeneous dielectric medium with an effective low frequency dielectric function $\epsilon_\text{eff}$, which is independent of $q$. In the presence of a nonhomogeneous dielectric background, the effective dielectric function, in general, exhibits a spatial dispersion~\cite{Profumo2012,SMB2012}, which, in turn, changes the $q$ dependence of the matrix elements of intra- and inter-layer Coulomb interaction. Thus, the dielectric background inhomogeneity can influence the plasmon dispersions but affect the electron polarization function. 
Notice also that the results reported here do not refer to structures of zigzag nanoribbons where the situation is qualitatively different~\cite{BF2007,Castro2009} because of the complexity of the single particle spectrum and of the respective Coulomb matrix elements, which cannot be represented only as functions of transferred momentum $q$.

\section{Conclusions}
In conclusion, the inter-band and intra-band contributions to the polarization function of Dirac fermions in metallic armchair graphene nanoribbons have been calculated and shown that the full polarization function exhibits a universal behavior, independent of the position of chemical potential and temperature. As a result, we find that individual and Coulomb-coupled graphene nanoribbons of a given width and spacing support fundamental plasmon modes with  energy dispersions, determined, within the random phase approximation, only by the graphene's fine structure constant.

\section{Acknowledgements}
The Center for Nanostructured Graphene (CNG) is sponsored by the Danish National Research Foundation, Project No. DNRF58. The work at the University of Antwerp was supported by the Flemisch Science Foundation (FWO-Vl) and the Methusalem Foundation of the Flemish Government. SMB gratefully acknowledges hospitality and support from the Department of Physics at the University of Missouri.

\appendix*

\begin{figure}[t]
\includegraphics[width=.9\columnwidth]{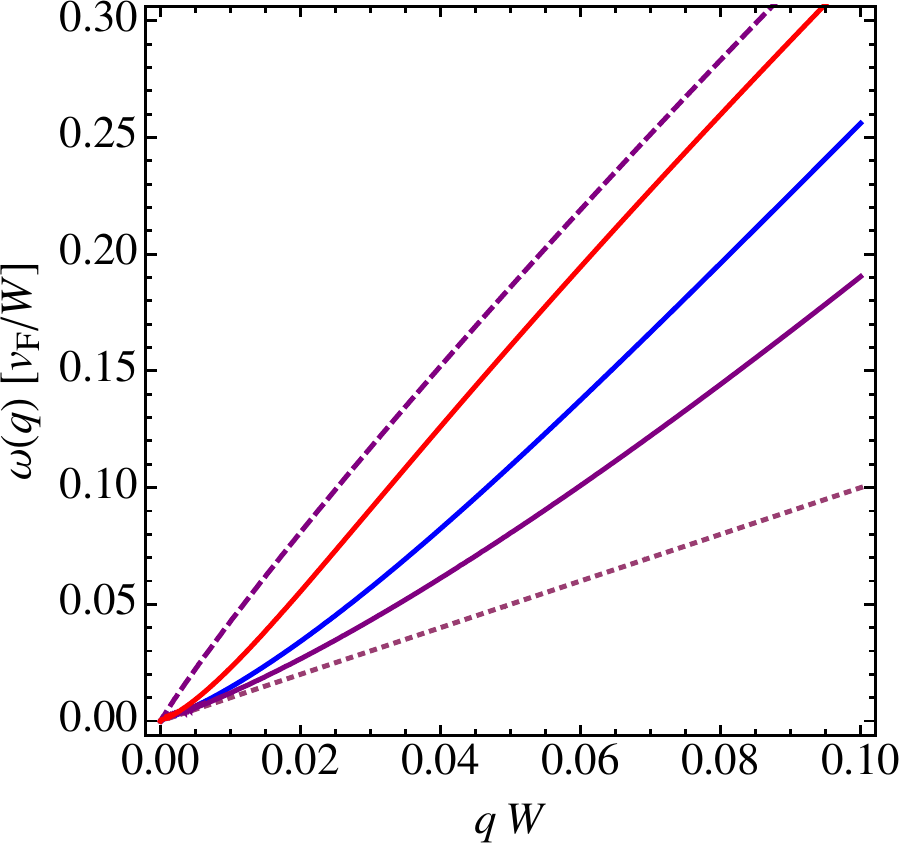}
\caption{Finite temperature effect on plasmons in individual metallic armchair nanoribbons. The solid curves from bottom to up correspond to the plasmon dispersions calculated, respectively, at temperatures $T_{1,2,3}=0.1, 0.04$, and $0.01v_{\text{F}}/W $. The dashed curve reproduces the plasmon dispersion calculated at $T=0$ in Ref.~\onlinecite{BF2007} and the dotted line corresponds to the single-particle Dirac spectrum.}
\label{fig4}
\end{figure}

\section{Inter-band plasmons at finite temperatures}

Here we demonstrate a strong temperature effect on the plasmon dispersions in metallic armchair graphene nanoribbons that appears if we restrict our consideration to inter-band electronic transitions ({\it cf.} Ref.~\onlinecite{BF2007}).

In this case the finite temperature effect is described by the function $F_{\text{inter}}(q,E_{\text{F}},T)$. Making use of its asymptote in Eq.~(\ref{phsinterq0}), the polarization function (\ref{pfinter}) in the long wavelength limit can be approximated as
\begin{equation}
\Pi_{\text{inter}}(q,\omega)\approx\frac{1}{2\pi v_{\text{gr}}}\frac{v_{\text{gr}}^{3} q^{3}}{T \left(\omega^2-v_{\text{gr}}^{2} q^2\right)}\left(1-\frac{E_\text{F}^2}{4 T^2}\right)~.
\label{piflwllapp}
\end{equation}
In this limit the Coulomb matrix element in individual nanoribbons are $v^{11}_{0;0}(q,0,W)\approx -2 \alpha_{gr} v_{\text{gr}}\ln{(q W)}$ and we find the plasmon dispersion $\omega_{p}(q) = c_{p}(q) q$, with the plasmon velocity given by
\begin{equation}
c_{p}(q) \approx v_{\text{gr}}\left[1-\frac{\alpha_{\text{gr}}}{2\pi} \frac{v_{\text{gr}}q }{T} \left(1-\frac{E_\text{F}^2}{4 T^2}\right)\ln(q W) \right]~.
\label{enlwll}
\end{equation}
As seen at finite temperatures the plasmon velocity remains finite in the limit of vanishing $q$. Accordingly, the energy dispersion of the plasmon lies close to the single-particle energy for small values of $q$ and deviates from it slowly with increasing $q$ at a speed of $-q^{2}\ln (q W)$.

In Fig.~\ref{fig4} we plot the energy dispersions of the plasmon versus $q$ in neutral nanoribbons, $E_{\text{F}}=0$, obtained numerically from Eqs.~(\ref{singscreening}), (\ref{pfinter}),  and (\ref{phsinter}). We calculate the plasmon energy at different, relatively high temperatures, $T_{1,2,3} W/v=0.1, 0.04$ and $ 0.01$, and compare them with the plasmon dispersion, calculated at $T=0$. It is seen that at finite temperatures the plasmon dispersions show qualitatively different behavior. For vanishing temperatures, $T\rightarrow 0$, the validity region of Eqs.~(\ref{piflwllapp}) and (\ref{enlwll}), $0<q<q_T$, shrinks a single point $q=0$, and the sequence of the solid curves in Fig.~\ref{fig4} tends smoothly to the dashed curve, obtained at zero temperature~\cite{BF2007}.


\begin{thebibliography}{99}
\bibitem{footnote} We note that for finite values of $E_{\text{F}}$ the formula in Eq.~(\ref{phsinter}) differs from the respective formula in Eq.~(15) from Ref.~\onlinecite{BF2007}.
\end{thebibliography}
\end{document}